\documentclass[%
 reprint,
 amsmath,amssymb,
 aps,
]{revtex4-2}

\usepackage[colorlinks,linkcolor=blue,urlcolor=blue,citecolor=blue]{hyperref}

\usepackage{subfigure}

\usepackage{graphicx}% Include figure files
\usepackage{dcolumn}% Align table columns on decimal point
\usepackage{bm}% bold math
\usepackage{graphicx}
\usepackage{booktabs}
\begin{document}

\preprint{APS/123-QED}

\title{Attosecond two-color x-ray free-electron lasers with dual chirp-taper configuration and bunching inheritance}% Force line breaks with \\Manuscript Title:\\with Forced Linebreak

\author{Hao Sun\textsuperscript{1}}
\author{Xiaofan Wang\textsuperscript{1}}
 \email{wangxf@mail.iasf.ac.cn}
\author{Weiqing Zhang\textsuperscript{2}}%
 \email{weiqingzhang@dicp.ac.cn}


\affiliation{
 1. Institute of Advanced Science Facilities, Shenzhen 518107, China\\
 2. Dalian Institute of Chemical Physics, Chinese Academy of Sciences, Dalian 116023, China\\
}%

\date{\today}% It is always \today, today,

\begin{abstract}
Attosecond x-ray pulses play a crucial role in the study of ultrafast phenomena occurring within inner and valence electrons. To achieve attosecond time-resolution studies and gain control over electronic wavefunctions, it is crucial to develop techniques capable of generating and synchronizing two-color x-ray pulses at the attosecond scale. In this paper, we present a novel approach for generating attosecond pulse pairs using a dual chirp-taper free-electron laser with bunching inheritance. An electron beam with a sinusoidal energy chirp, introduced by the external seed laser, passes through the main undulator and afterburner, both with tapers. Two-color x-ray pulses are generated from the main undulator and afterburner, respectively, with temporal separations of hundreds of attoseconds to several femtoseconds and energy differences of tens of electron volts. Notably, the afterburner is much shorter than the main undulator due to bunching inheritance, which reduces the distance between two source points and alleviates the beamline focusing requirements of the two-color pulses. A comprehensive stability analysis is conducted in this paper, considering the individual effects of shot noise from self-amplified spontaneous emission and carrier-envelope phase jitter of the few-cycle laser. The results show that the radiation from the afterburner exhibits excellent stability in the proposed scheme, which is beneficial for x-ray pump-probe experiments. The proposed scheme opens up new possibilities for attosecond science enabled by x-ray attosecond pump-probe techniques and coherent control of ultrafast electronic wave packets in quantum systems.

\end{abstract}

%\begin{description}
%\end{description}

\maketitle

%\tableofcontents

\section{\label{sec:level1}INTRODUCTION}
% \protect\\ The line
 X-ray free-electron lasers (XFELs) \cite{McNeil2010,PhysRevSTAB.10.034801} have enabled scientists to explore dynamic processes on the femtosecond time scale by taking advantage of the short pulse duration and the high intensity of FEL pulses. In time-resolved studies, pump-probe experiments are usually realized by combining conventional laser pumping with an x-ray FEL probe. This technique allows monitoring the temporal evolution of various photo-induced processes on ultrafast time scales \cite{kraus_2023_kc9ry-qb847}. However, a significant challenge in these experiments is the inherent temporal jitter between the pump and probe pulses arising from the independent laser sources \cite{femo}. To address this issue, researchers have developed new FEL sources capable of generating dual pulses with different photon energies from the same electron beam \cite{PhysRevSTAB.13.030701}. This advancement has facilitated x-ray pumping and x-ray probing experiments while reducing the temporal jitter between the two pulses \cite{lcls}. Nevertheless, despite the progress made in these femtosecond time-resolved experiments, they have been primarily limited to observing atomic motion. The motion of electrons, which occurs on timescales of hundreds of attoseconds, remains elusive. Moreover, in addition to the observation of ultrafast processes, a longstanding goal in atomic physics has been to manipulate electronic wavefunctions with attosecond precision. To achieve attosecond time resolution and gain control over electronic wavefunctions, it is crucial to develop techniques capable of generating and synchronizing two-color x-ray pulses at the attosecond scale. 
 
 At the femtosecond time scale, some of the proposed two-color FEL methods have been experimentally demonstrated and successfully applied to scientific experiments. In general, two-color FEL pulses can be obtained by tuning the resonance of the undulator or by manipulating the electron beam properties. For the first approach, the undulator is divided into two parts, each of which resonates at a different wavelength \cite{Hara2013}. The temporal separation of the two-color FEL pulses can be controlled using a magnetic chicane between the undulator modules. However, this method requires the same electrons to generate both two-color pulses, resulting in FEL operation far from saturation. Alternatively, several fresh-slice-based methods have been proposed to generate two-color FEL pulses, relying on different parts of the electron bunches lasing in different undulators \cite{freshslice1,freshslice2,freshslice3,freshslice4,PhysRevLett.110.134801}. Another type, based on the manipulation of the electron beam, involves allowing electrons with different energies to generate two-color FEL pulses in the same undulator. Several methods of electron beam manipulation have been proposed, including the generation of twin bunches within the same RF bucket \cite{Marinelli2015}, the creation of two current spikes through wake field effects \cite{wake}, the use of a double-slotted foil within the beam as an emittance spoiler \cite{slotted}, nonlinear electron compression \cite{nonlinear2}, and a photocathode laser emittance spoiler \cite{laser_emitt}. Furthermore, the laser-based seeded FELs have been experimentally demonstrated to deliver two synchronized pulses of different colors for novel pump-probe experiments \cite{seeded_two_color}.

Attosecond soft x-ray pulses play a vital role in comprehending the rapid quantum mechanical motion of electrons in molecular systems. Historically, high harmonic generation (HHG) sources have been the cutting-edge approach for generating attosecond pulses \cite{HHG1,HHG2,HHG3,HHG4,HHG5}. However, their pulse energy output does not exhibit a favorable scaling relationship with the photon energy \cite{HHG5}.  Attosecond science is rapidly advancing within the FEL community, thanks to various proposed schemes such as the chirp-taper scheme \cite{chirp_taper1,chirp_taper2,chirp_taper3,chirp_taper4,chirp_taper5,chirp_taper6,chirp_taper7}, enhanced self-amplified spontaneous emission scheme (ESASE) \cite{ESASE_1,ESASE_2,ESASE_3,ESASE_4,ESASE_5,ESASE_6,ESASE_7,ESASE_8,ESASE_9}, and seeded FEL schemes \cite{PhysRevSTAB.12.060701,Xiao}.  To conduct attosecond pump-attosecond probe experiments effectively, it is necessary to generate synchronized pulse pairs. Ideally, these pulse pairs should possess distinct photon energies, enabling the investigation of excitation at one atomic site in a molecular system while probing another site. Therfore, a method was proposed to generate pairs of sub-femtosecond x-ray pulses with attosecond time stability and controllable separation. This method relies on a double chirp-taper scheme driven by self-modulation in a magnetic wiggler \cite{chirp_taper3}. Recently, another scheme has also emerged, aiming to produce sub-femtosecond pulse pairs with femtosecond-scale time separation and few-electron-volt energy separation based on frequency pulling \cite{robles:ipac23-tupl094}.

In this paper, we present a novel approach for generating two-color attosecond pulses using a dual chirp-taper free-electron laser with bunching inheritance.  Initially, an electron beam undergoes significant energy modulation through interaction with a few-cycle laser in a wiggler, resulting in the creation of a sinusoidal energy-chirp phase space that comprises a main positive energy chirp part and a negative energy chirp part. The energy-modulated electron beam then passes through a main undulator with an appropriately designed taper. If the energy modulation exceeds the FEL energy bandwidth and the chirp-taper matching condition is satisfied, lasing occurs in a short fraction of the positive energy chirp part, which leads to the generation of ultra-short FEL pulses \cite{chirp_taper1,chirp_taper2,chirp_taper3,chirp_taper4,chirp_taper5,chirp_taper6,chirp_taper7}. Next, the electron beam is directed into a magnetic chicane, which introduces a precise time delay of a few femtoseconds, allowing for the separation of the first FEL pulse. Importantly, the bunching of the electron beam retains a high value, facilitating its reuse for subsequent FEL processes. Subsequently, the electron beam traverses a short, tapered afterburner, which is tailored to match the negative energy chirp. Driven by the bunching formed in the main undulator, the electron beam generates an intense FEL pulse in a short afterburner. The wavelength of the second FEL pulse undergoes a blue shift due to the compression of the bunching in the afterburner. As a result, the photon energy of the second FEL pulse differs from that of the first FEL pulse. In general, when the distance between the two-color source points is too large, it often prevents one of the lights from being effectively focused at the sample position \cite{EURO_two_color}. Notably, the reuse of microbunching enables a significantly shorter afterburner length compared to the main undulator. This advantage greatly reduces the distance between the two-color source points, thereby easing the requirements for beamline focusing. Furthermore, the power and time stability of the FEL pulses from the afterburner are improved due to the reuse of the bunching.

The remainder of the paper is organized as follows: In Sec~\ref{sec2}, we provide an overview of the principles behind the proposed scheme. In Sec~\ref{sec3}, we will present the single-shot simulation results to demonstrate our scheme. Furthermore, we will present the stability analysis, considering shot noise from self-amplified spontaneous emission and the carrier-envelope phase of the few-cycle laser, to demonstrate the reliability of the proposed scheme. Finally, we will provide discussions and a brief summary of this work in Sec~\ref{sec4}.

\begin{figure*} 
\includegraphics[width=15cm]{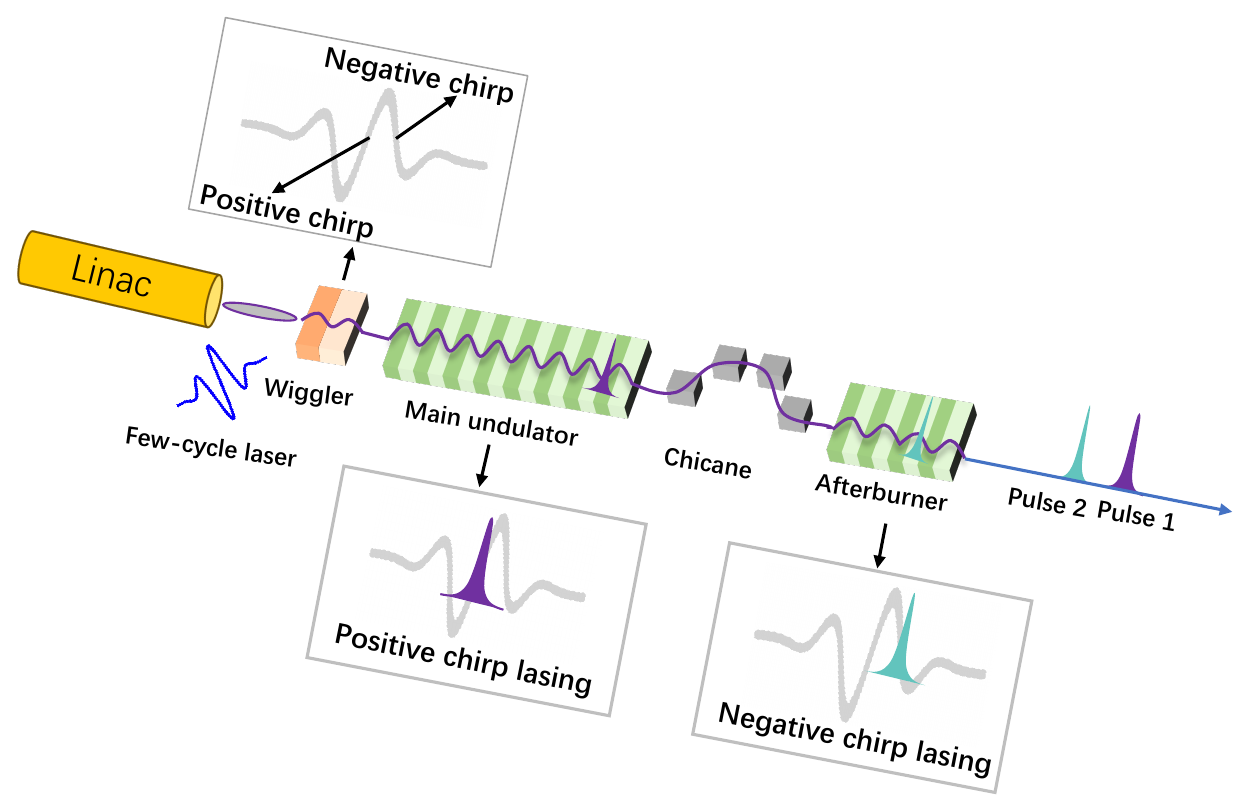}% Here is how to import EPS art[b]
\caption{ Schematic of the proposed scheme for generating attosecond two-color FEL pulses. }
\label{figure1}
\end{figure*}

\section{PRINCIPLE} \label{sec2}
The schematic of the proposed scheme is shown in Fig.~\ref{figure1}. A few-cycle laser at 800 nm wavelength interacts strongly with the central part of an electron bunch in a two-period wiggler, resulting in a region with a sinusoidal energy modulation. As shown in Fig.~\ref{figure1}, this region comprises a main positive energy chirp part and a negative energy chirp part. The energy-modulated beam then enters a long tapered undulator, which is called the main undulator, for FEL processes. According to the principle of the chirp-taper scheme \cite{chirp_taper1,chirp_taper2,chirp_taper3,chirp_taper4,chirp_taper5,chirp_taper6,chirp_taper7}, if both the positive energy chirp and the main undulator taper are properly matched, a slice with the strongest positive energy chirp is selected for lasing. The lasing in the rest of the bunch is strongly suppressed as these electrons fall out of resonance. To achieve precise control over the temporal delay between the double pulses with femtosecond precision, a chicane is positioned between the main undulator and the afterburner. This chicane can be utilized to delay the electron beam relative to the first FEL pulse. It should be noted that the peak bunching of the electron beam decreases as it passes through the chicane. However, if the longitudinal dispersion of the chicane is appropriately designed, the bunching of the electron beam can still maintain a significant value, allowing for reuse in the afterburner. Finally, the electron beam will travel through a short, tapered afterburner, which is matched to the negative energy chirp. Driven by the bunching formed in the main undulator, the electron beam can produce an intense FEL pulse in the afterburner. The wavelength of the FEL pulse generated by the afterburner will blue-shift due to the compression of the bunching when the negative energy chirp is compensated by the undulator taper \cite{chirp_taper5}. Therefore, the two-color FEL pulses can be generated in the proposed scheme.

\begin{table}%[b]%The best place to locate the table environment is directly after its first reference in text
\caption{\label{table1}%
Parameters of the electron beam and undulators used in the simulations.
}
\begin{ruledtabular}
\begin{tabular}{lcdr}
\textrm{Parameter}&
\textrm{Value}\\
%\multicolumn{1}{c}{\textrm{Decimal}}&
%\textrm{Right}\\
\colrule
Energy  & 2.53 GeV  \\
Peak current & 2 kA \\
Emittance & 0.45 mm$\cdot$mrad  \\
Uncorrelated energy spread  & 130 keV \\
Wiggler/Main undulator/Afterburner period & 16/3/3 cm \\
Wiggler/Main undulator/Afterburner length & 0.32/32/4 m \\

\end{tabular}
\end{ruledtabular}
\end{table}

\section{Simulations} \label{sec3}
To investigate the performance of the proposed scheme, three-dimensional simulations were performed with the parameters of a typical soft x-ray free-electron laser facility, as listed in Table \ref{table1}. The electron beam extracted from a linear accelerator was characterized by the following parameters: an energy of 2.53 GeV, a peak current of 2 kA, rms normalized emittance of 0.45~mm$\cdot$mrad, and an uncorrelated energy spread of 130 keV. For our simulations, we employed a 1.5-cycle laser with a carrier wavelength of 800 nm and a pulse energy of 0.177 mJ, which is currently available with state-of-the-art laser technology \cite{few_cycle}. The wiggler has two periods, with a period length of 16 cm and a K value of 22.
To simulate the energy modulation in the wiggler, we used the FALCON software \cite{FALCON}, which is built based on the fundamental electrodynamic theory. The FEL simulations were performed using the three-dimensional time-dependent code Genesis 1.3 \cite{reiche1999genesis}. 

\subsection{Single-shot simulation}
\begin{figure}
\includegraphics[width=1\columnwidth]{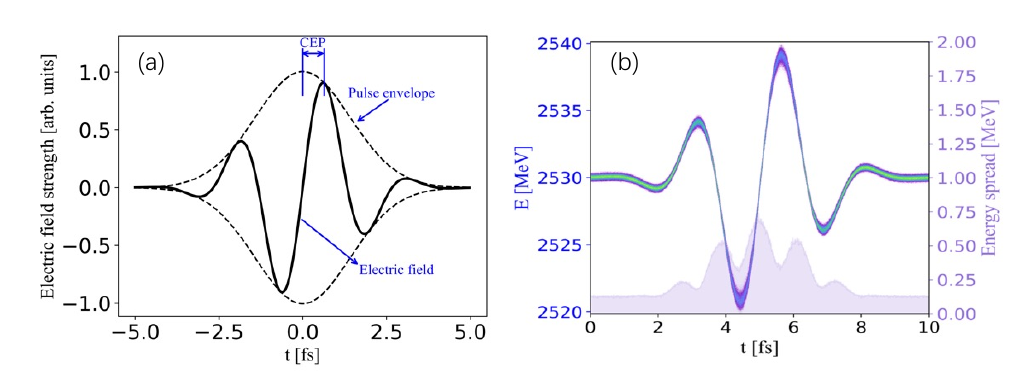}% Here is how to import EPS art[b]
\caption{(a) Normalized electric field of the few-cycle laser pulse. (The pulse intensity envelope is shown by the dotted line and the oscillating electric field is shown by the solid line.) (b) The longitudinal phase space of the electron beam after the wiggler. Bunch head is on the right side.}
\label{figure2}
\end{figure}

In the following section, we will illustrate the proposed scheme with numerical simulations utilizing the above parameters. The electron beam will achieve an energy modulation of order 10 MeV by interacting with a few-cycle laser in a wiggler, and the electric field of the laser pulse normalized to the peak value is depicted in Fig.~\ref{figure2}(a). The longitudinal phase space of the modulated electron beam is presented in Fig.~\ref{figure2}(b). It can be shown that a sinusoidal energy modulation with an 800 nm period and an energy modulation amplitude of order 10 MeV is imprinted into the electron beam. The maximum slope of the positive energy chirp can be estimated by the energy modulation amplitude $\Delta \gamma$ and the modulation period $\lambda_m$

\begin{equation}
\left|\frac{d \gamma}{d s}\right|=\frac{2 \pi}{\lambda_m} \Delta \gamma .
\label{eq1}
\end{equation}

\begin{figure}
\includegraphics[width=0.8\columnwidth]{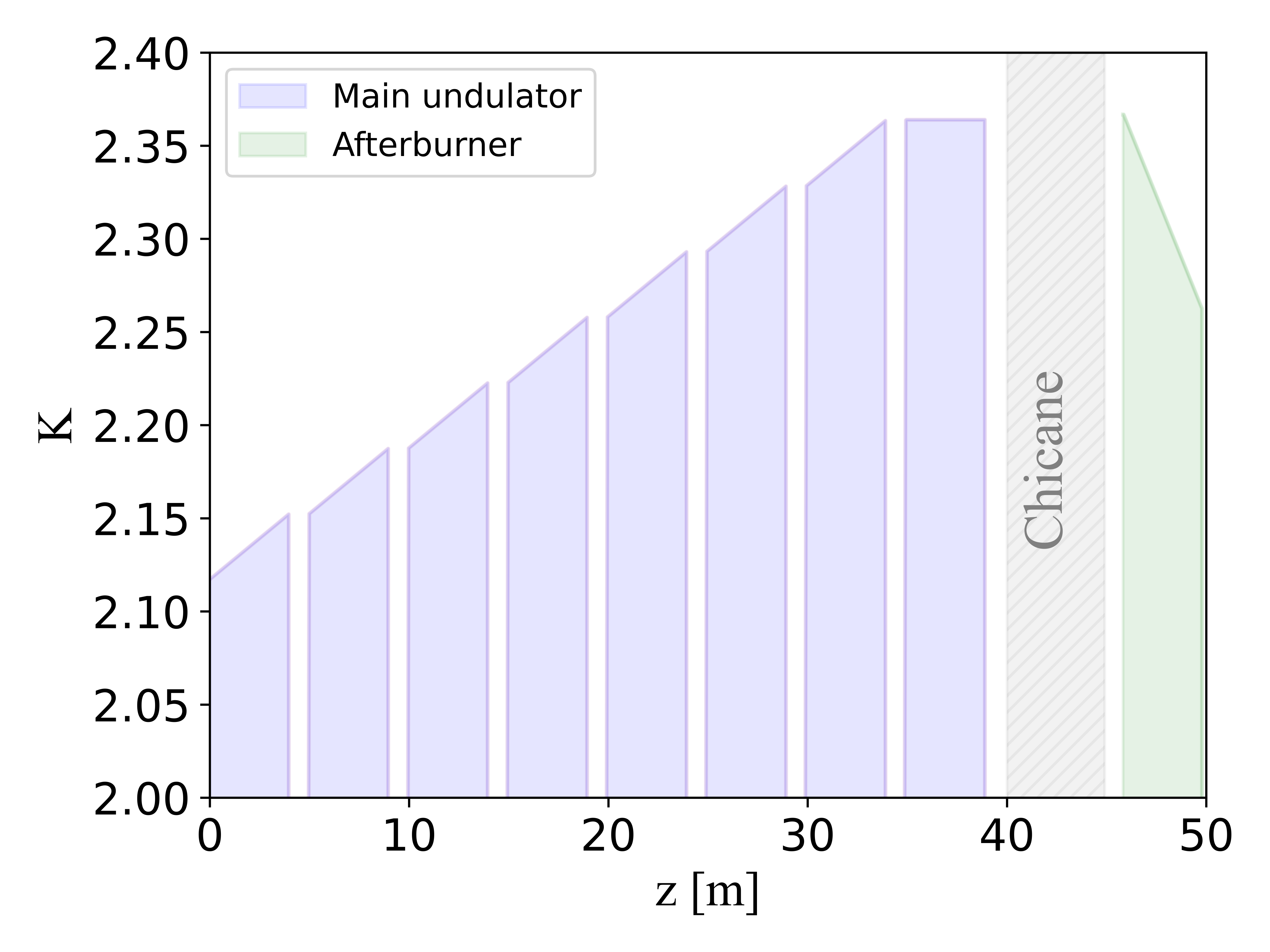}% Here is how to import EPS art[b]
\caption{The optimized undulator taper of the main undulator and the afterburner.}
\label{figure3}
\end{figure}

It has been demonstrated that a gain degradation of a linear energy chirp can be compensated by a linear undulator taper when the following condition is satisfied \cite{chirp_taper2}:

\begin{equation}
\frac{d K}{d z}=\frac{\left(1+K_0^2 / 2\right)^2}{K_0} \frac{1}{\gamma_0^3} \frac{d \gamma}{c d t}.
\label{eq2}
\end{equation}
Here, $K$ is the undulator parameter, ${K_0}$ is the value of the undulator parameter at the main undulator entrance, ${\gamma_0}$ is the relativistic factor for a reference particle, and $c$ is the speed of light. For the energy-modulated electron beam as shown in Fig.~\ref{figure2}, the taper strength is estimated according to Eq.~\ref{eq2}, and then optimized to generate an isolated attosecond FEL pulse. The main undulator consists of eight 4-meter-long variable-gap undulator segments with quadrupoles in the intersections. In our simulation, the undulator is tuned to 2 nm at the entrance ($K_0=2.12$), and the undulator parameter $K$ increases by 0.0352 in each undulator segment, as illustrated in Fig.~\ref{figure3}. The afterburner has an opposite taper compared with the main undulator and we will illustrate it in the next paragraph. 

\begin{figure}
\includegraphics[width=1\columnwidth]{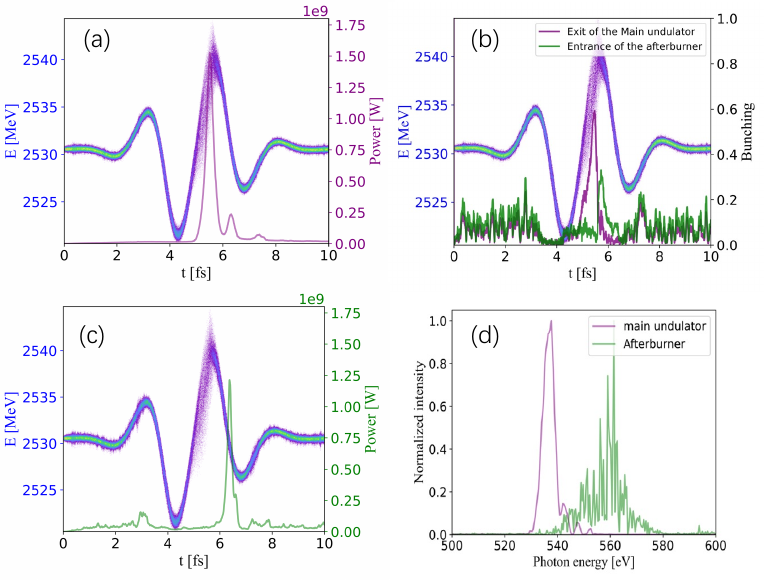}% Here is how to import EPS art[b]
\caption{(a) Energy phase space of the electron beam and a profile of the radiation pulse (purple line) at the exit of the main undulator. (b) Bunching profiles at the exit of the main undulator (purple line) and at the entrance of the afterburner (green line). (c) Energy phase space of the electron beam and a profile of the radiation pulse (purple line) at the exit of the afterburner. (d) Spectra of the pulses from the main undulator (purple line) and the afterburner (green line).}
\label{figure4}
\end{figure}

According to the optimization value of the taper, simulation results indicate that lasing will occur on a short fraction of the electron beam, leading to the generation of an isolated attosecond FEL pulse with a peak power of 1.5 GW and a FWHM pulse length of 317 as, as presented in Fig.~\ref{figure4}(a). In our proposed scheme, a small chicane is used to delay the electron beam so that the first pulse from the main undulator and the second pulse from the afterburner to be longitudinally separated. This not only introduces a small dispersion, characterized by a transfer matrix coefficient $\mathrm{R} _{56}=0.9 \mu \mathrm{m}$ (a time delay of 1.5 fs), but also affects the bunching of electron beams, where the maximum bunching factor changes from 0.6 to 0.3 and the width of the bunching is reduced, as depicted in Fig.~\ref{figure4}(b). It should be noted that the bunching of 0.3 can still be reused for lasing in the next stage. Driven by the bunching formed in the main undulator, the electron beam can produce intense FEL pulses in the afterburner. Finally, the electron beam passes through a tapered afterburner (one 4-meter-long undulator segment), which is matched to the negative energy chirp. The undulator parameter of the afterburner is shown in Fig.~\ref{figure3}, with the value decreasing along the afterburner.

\begin{figure}
\includegraphics[width=1\columnwidth]{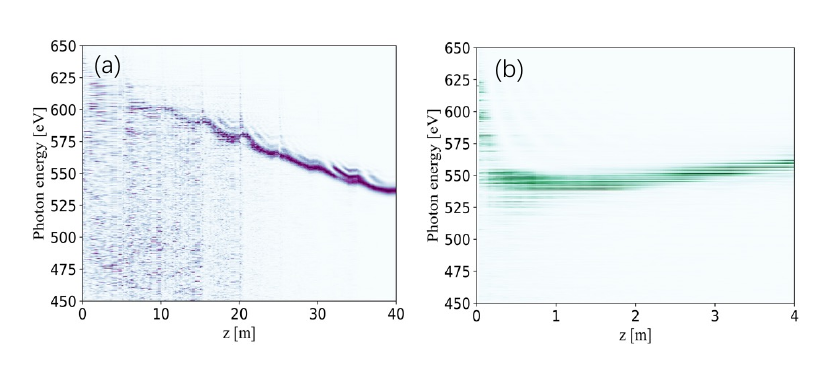}% Here is how to import EPS art[b]
\caption{FEL spectra evolution along the main undulator (a) and the afterburner (b).}
\label{figure5}
\end{figure}

As shown in Fig.~\ref{figure4}(c), an isolated pulse with a peak power of 1.2 GW and a FWHM pulse length of 198 as can be generated in the afterburner. It can be seen that pulse length in the afterburner is reduced due to the narrowing of the bunching width as shown in Fig.~\ref{figure4}(b). The spectra of the pulses from the main undulator and the afterburner are presented in Fig.~\ref{figure4}(d). It is apparent that photon energies of the two pulses are different, with a difference in photon energy of about 30 eV.  Fig.~\ref{figure5} shows the FEL spectra evolution along the main undulator and the afterburner. It can be seen that the FEL spectra from the main undulator and the afterburner are shifted in opposite directions. The wavelength shifting of the FEL pulses is due to the compression or decompression of the bunching when the energy chirp is compensated by the undulator taper. The compression or decompression of the FEL wavelength can be estimated as \cite{chirp_taper5}

\begin{equation}
C=\frac{1}{1 \pm R_{56} \frac{d \gamma}{c \gamma_0 d t}},
\label{eq3}
\end{equation}
where $R_{56}$ is the momentum compression factor, and the $\pm$ represents compression or decompression. After bunching compression or decompression, the radiation wavelength undergoes a transformation from $\lambda_1$ to $\lambda_2=\lambda_1$/C. In the case of the main undulator, $R_{56}$ mainly comes from the dispersion of the undulator $2 \lambda N_{u n d}$, where $N_{u n d}$ is a number of periods of the undulator, and $\gamma$ is the resonance wavelength. In the case of the afterburner, $R_{56}$ comprises two components. The first component arises from the dispersion introduced by the chicane, while the second component originates from the undulator. According to Eq.~\ref{eq3}, the photon energy difference between the first pulse and the second pulse is approximately 27 eV, which roughly matches the simulation results depicted in Fig.~\ref{figure5}. Due to the continuous shifting of the photon energy in the afterburner, as shown in Fig.~\ref{figure5}(b), a fine adjustment of the energy difference between two-color pulses can be achieved by tuning the length of the afterburner.

\begin{figure}
\includegraphics[width=0.7\columnwidth]{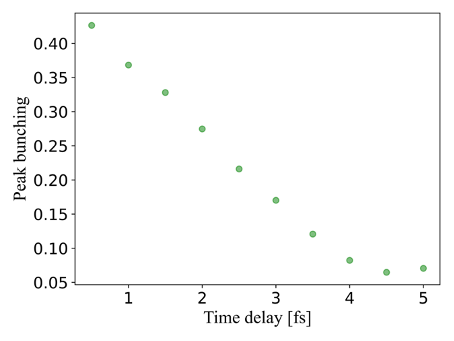}% Here is how to import EPS art[b]
\caption{Variation of the peak bunching with the time delay of the chicane.}
\label{figure6}
\end{figure}

In addition, precise adjustment of the time delay between the two pulses is crucial for two-color x-ray pump-probe experiments. Our proposed approach involves utilizing the bunching formed in the main undulator for secondary FEL emission in the afterburner. However, if the delay provided by the chicane (indicated by a large $R_{56}$ value) is excessively long, it results in a reduction of the bunching, thereby impacting the effectiveness of the secondary FEL emission. In the subsequent research, we have examined the relationship between peak bunching and time delay to determine the achievable range of temporal tuning with this approach. As shown in Fig.~\ref{figure6}, the bunching decreases as the time delay increases after passing through the chicane. However, it is notable that within a time delay range of 4 fs, the bunching still maintains a significant value, enabling it to drive FEL radiation in the afterburner. Nevertheless, this may lead to a decrease in FEL output power, which can be improved by increasing the length of the afterburner. Additionally, to efficiently control $R_{56}$ while achieving a longer time delay, a chicane system equipped with quadrupole magnets in the dispersion region may offer a more flexible solution \cite{thompson:fel19-thp033}.

\subsection{Muti-shot simulations}

\begin{figure}
\includegraphics[width=1\columnwidth]{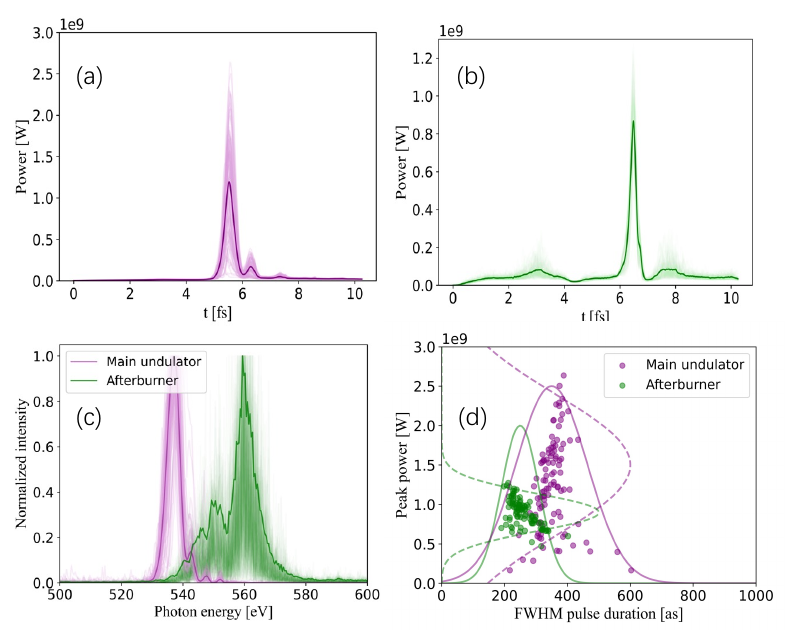}% Here is how to import EPS art[b]
\caption{(a) Power profiles of the pulses from the main undulator. The purple line represents the average power profile of 100 shots. (b) Power profiles of the pulses from the afterburner. The green line represents the average power profile of 100 shots. (c) Spectra of the two-color pulses. (d) Statistics plots of the peak power and pulse duration of the 100-shot two-color pulses.}
\label{figure7}
\end{figure}

To evaluate the stability of the proposed scheme, we conducted multi-shot simulations with different shot noise initializations. Specifically, we performed 100-shot simulations to study the properties of two-color attosecond pulses. Fig.~\ref{figure7}(a) and Fig.~\ref{figure7}(b) illustrate the corresponding 100-shot power profiles and the average power profile, respectively, generated by the main undulator and the afterburner. Additionally, Fig.~\ref{figure7}(c) presents the spectrum profiles for the 100 shots as well as the average spectrum profile. Furthermore, statistical results regarding the pulse duration and peak power are illustrated in Fig.~\ref{figure7}(d), where Gaussian fits are employed to represent the distributions of pulse duration and peak power across the 100 shots. The standard deviation of the pulse duration is found to be 113 as in the main undulator and 59 as in the afterburner. Additionally, the standard deviation of the peak power is 0.9 GW in the main undulator and 0.3 GW in the afterburner, as depicted in Fig.~\ref{figure7}(d). This indicates that the stability of the FEL pulses from the afterburner is significantly improved compared to those from the main undulator. The stability arises from the adjustment of the bunching formed by the self-amplified spontaneous emission process in the main undulator by the chicane, whereby lower bunching is increased and higher bunching is reduced for different shot-noise initializations.

\begin{figure}
\includegraphics[width=1\columnwidth]{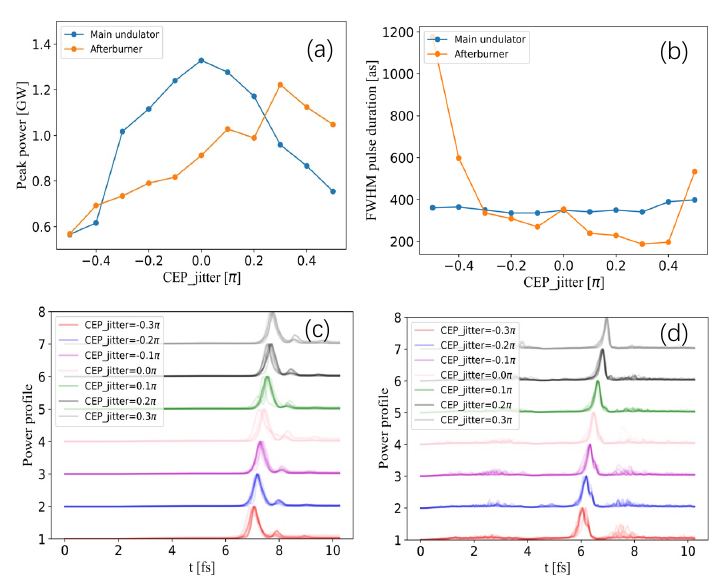}% Here is how to import EPS art[b]
\caption{The stability analysis of the peak power (a) and pulse duration (b) from the main undulator and the afterburner in relation to the CEP phase jitter (each point represents the average value of ten shots). Power profiles of the pulses from the main undulator (c) and the afterburner (d) under different CEP phase jitter conditions.}
\label{figure8}
\end{figure}

The carrier-envelope phase (CEP) of a few-cycle laser pulse refers to the relative timing between the carrier wave and the envelope of the laser pulse. It plays a crucial role in the properties and applications of ultrafast laser systems. In particular, the CEP stability of the 1.5-cycle laser used in our scheme is crucial for the structure of the electron beam modulation generated by the laser, and thus on the structure of the emitted x-rays. It ensures reproducibility and accuracy in the proposed scheme, enabling long-term experiments. Therefore, a stability analysis of the proposed scheme with respect to the CEP of the few-cycle laser is presented in Fig.~\ref{figure8}. Fig.~\ref{figure8}(a) and Fig.~\ref{figure8}(b) represent the stability analysis of the peak power and the pulse duration from the main undulator and the afterburner in relation to the CEP phase jitter, respectively. Fig.~\ref{figure8}(c) and Fig.~\ref{figure8}(d) illustrate the power profiles of the pulses from the main undulator and the afterburner for the different CEP phase jitters, respectively. It can be observed that the two-color FEL pulses can be generated stably within a $\pm 0.3 \pi$ phase change of the CEP, which is a reasonable value achievable with current laser technology.

\section{CONCLUSIONS AND DISCUSSIONS}
\label{sec4}
In this paper, we present a novel approach for generating attosecond pulse pairs using a dual chirp-taper free-electron laser with bunching inheritance. Simulations demonstrate that the proposed scheme is capable of generating GW-level two-color attosecond x-ray pulses with temporal separations ranging from hundreds of attoseconds to several femtoseconds and energy differences in the tens of electron volts. In addition, the afterburner can be considerably shorter than the main undulator due to the reusing of the microbunching, thereby reducing the distance between the two-color source points and alleviating the beamline focusing requirements for the two-color pulses. Furthermore, we also conduct stability analysis, considering shot noise from self-amplified spontaneous emission and the carrier-envelope phase of the few-cycle laser, to demonstrate the reliability of the proposed scheme. The simulation results indicate that the radiation from the afterburner exhibits excellent stability, which is more advantageous for attosecond x-ray pump-probe experiments. 

The proposed scheme is an extension of previous work on chirp-taper FEL \cite{chirp_taper1,chirp_taper2,chirp_taper3,chirp_taper4,chirp_taper5,chirp_taper6,chirp_taper7}. Compared to the modified chirp-taper scheme \cite{chirp_taper5}, this approach further enhances the performance of generating two-color FEL pulses. Unlike the double chirp-taper scheme \cite{chirp_taper3}, two-color generation in the proposed scheme is based on compressing and decompressing the bunching within the undulator rather than completely wiping out the bunching and resonating it to a new wavelength. The reuse of the bunching enables a significantly shorter length of the afterburner. In this approach, the energy difference between the two-color pulses is on the order of tens of electron volts. However, it is worth noting that the afterburner in the proposed scheme can also serve as a harmonic afterburner, further increasing the energy difference between the two-color pulses. Therefore, this approach serves as a valuable complement to the previous chirp-taper, two-color FEL scheme \cite{chirp_taper3}. The proposed scheme provides new possibilities for attosecond science enabled by x-ray attosecond pump-probe techniques and coherent control of ultrafast electronic wave packets in quantum systems.

\begin{acknowledgments}
The authors would like to thank Fucai Zhang (SUSTech), Wei Liu (IASF) and Yinpeng Zhong (IASF) for useful suggestions and comments on this work. The authors also would like to express their gratitude to Li Zeng, Lingjun Tu, Huaiqian Yi, Yifan Liang, and Yong Yu for their fruitful discussions in FEL physics and simulations. This work is supported by the Shenzhen Science and Technology Program (Grant No. RCBS20210609104332002), the Scientific Instrument Developing Project of Chinese Academy of Sciences (Grant No. GJJSTD20220001), and the National Natural Science Foundation of China (Grant No. 22288201).
\end{acknowledgments}

\nocite{*}

\end{document}